\documentclass[twocolumn,showpacs,prl,amsmath,amssymb]{revtex4}
\usepackage{graphicx}
\usepackage{dcolumn}
\usepackage{bm}

\begin{document}

\preprint{APS/123-QED}

\title{Porosity dependence of sound propagation in liquid $^4$He filled aerogel}

\author{K. Matsumoto$^1$, Y. Matsuyama$^1$, D. A. Tayurskii$^{1,2}$, and K. Tajiri$^3$}
\affiliation{
$^{1}$Department of Physics, Kanazawa University, Kakuma-machi, Kanazawa 920-1192, Japan\\
$^{2}$Physics Department, Kazan State University, Kremlevskaya str., 18, Kazan, 420008, Russia\\
$^{3}$Ceramics Research Institute, National Institute of Advanced Industrial Science and Technology, Nagoya 463-8560, Japan}

\date{\today}

\begin{abstract}
Longitudinal sound wave propagation has been studied in an aerogel-liquid $^4$He system for various porosities of aerogel.
The superfluid transition was identified as the absorption peak, whose magnitude was suppressed by aerogel.
The sound velocity was analyzed within a hydrodynamic theory in both normal and superfluid phases.
The absorption peak due to phonon-roton interaction around 1 K was not observed even with the most porous aerogel.
The low temperature sound velocity and attenuation show that direct collisions of phonons with aerogel strands plays an important role in the acoustic properties.

\end{abstract}

\maketitle

%\section{INTRODUCTION}
%Introduction%
%background%
Porous media filled with fluid have been intensively studied
experimentally and theoretically because of their physical and
technological importance. The effect of disordered pore
structures on the properties of the fluid can be examined in these
systems. 
There has been considerable interest
in the behavior of superfluid $^4$He in the presence of a random
disorder induced by highly open porous media. Recent experiments
on the superfluid transition of $^4$He contained in porous media
such as Aerogel, Xerogel and Vycor glass have revealed that
the superfluid transition differs from that of bulk $^4$He
\cite{Wong,Chan}. 
The superfluid transition of $^4$He in aerogel has been observed to be sharp\cite{Wong,Chan}, and has suggested 
to manifest a genuine phase transition.
The transition temperature in aerogel T$_c$ has been suppressed with decreasing aerogel porosity. 

Understanding the results of acoustic experiments is important when dealing with porous media. 
Use of liquid $^4$He offers unique advantages due to the existence of the superfluid phase with more than one
sound mode. 
The bulk fluid displays two propagating modes: first sound (a compressional wave) and second sound 
(a temperature wave)\cite{Khalatnikov}. 
In a porous media where the normal component is 
clamped by its viscosity and only the superfluid component
can move, fourth sound (relative motion of the superfluid and normal fluids) propagates and can be used to determine the
superfluid fraction.

Longitudinal and transverse ultrasound velocity have been measured
in $^4$He filled Vycor glass\cite{BeamishHikata}. Warner and
Beamish\cite{WarnerBeamish} studied the transverse sound velocity 
and attenuation in alumina ceramics with various porosities. 
They argued that the experimental results in both the low and high frequency regimes
for normal and superfluid phases can be quantitatively elucidated
by the Biot model\cite{Biot1,Biot2,Biot3,Biot4}.

%aerogel%
Silica aerogels are synthesized via a sol-gel process and
hypercritical drying which enable production of tenuous solids with porosity $\phi$ 
as large as 99.8 $\%$ and unique acoustic properties. 
Silica aerogel is thought of as a network of nanoscale SiO$_2$ strands. 
The elastic moduli of aerogels are a few orders of magnitude smaller than that of bulk solids 
and the sound speed substantially depends on the porosity. 
Ultrasound measurements have shown sound speeds as low as 20 m/s for the highest porosity aerogel\cite{Fricke}. 

The high-porosity aerogels are so soft that the aerogel matrix and the clamped normal fluid moves as the results of pressure and temperature gradients, unlike other porous media.
%McKenna%
This results in sound modes intermediate between first and fourth sound\cite{MaKenna1} and a second-sound-like mode\cite{MaKenna2}.
McKenna et al. \cite{MaKenna2} calculated the longitudinal sound velocity for the two modes using modified two-fluid hydrodynamic equations in order to take aerogel motion into consideration.
They also observed the propagation of both the fast (intermediate between first and fourth sound) and the slow (second-sound-like) modes in $^4$He in aerogel from 1.1 K to T$_c$.
They found agreement of the model with the observed sound velocity within the experimental temperature range.

%Motivation%
We have observed the 10 MHz longitudinal ultrasound signal with three
different porosity (92.6, 94.0 and 94.8 \%) aerogels from 0.5 to 4.2 K, and measured the
sound velocity and attenuation in order to study sound propagation
in the liquid $^4$He filled aerogel system both in the normal and
superfluid phases. 
Preliminary results have been published
elsewhere\cite{Matsumoto12}. 
The viscous penetration depth of
liquid $^4$He at 10 MHz (about 40 nm) is estimated to be longer than the typical
SiO$_2$ strand distance (about 10 nm for 98 \% aerogel and much shorter for 95 \%) ; thus the normal fluid in these systems
is expected to be completely locked to aerogel matrix by
viscosity. 
The sound velocity of aerogel largely depends on the
porosity so that we can obtain an aerogel whose sound velocity is
larger or smaller than that of bulk fluid. 
%It is interesting to see what would happen if the relation of sound velocities between aerogel and fluid is reversed.
Sound experiment with aerogel has an advantage to investigate what happens when the relation of sound velocities between fluid and porous media is changed. 

%experiment%
%\section{EXPERIMENT}
Three aerogels were grown by a sol-gel process from tetramethoxysilane (TMOS) as a one step process. 
The porosities were determined using a standard dry weight method. 
We machined aerogels into cylinders (7 mm in diameter) which were enclosed in brass shells of 8.5 mm diameter and 3.0 mm length. 
The ends of the samples were polished flat and parallel. 
The sample cylinder was sandwiched between two LiNbO$_{3}$ transducers with springs. 
We used the same pair of transducers for every sample in order to easily compare the signal attenuation among aerogels. 
Aerogels were immersed in liquid $^{4}$He at SVP. 
Temperatures were measured with a ruthenium oxide resistance thermometer and stabilized using a PI controller.
The ultrasonic measurements were made using a standard pulse transmission and a phase sensitive detection technique.

%Results and discussion%
%\section{RESULTS AND DISCUSSION}
%overall%
We were able to observe the sound signal throughout the temperature range from 0.5 to 4.2 K. 
The transmitted sound signal through the aerogel in a vacuum, however, could not be observed because of the imperfect connection between the transducers and aerogel or large attenuation from the aerogel itself.

\begin{figure}[htb]
\includegraphics[width=6cm,keepaspectratio]{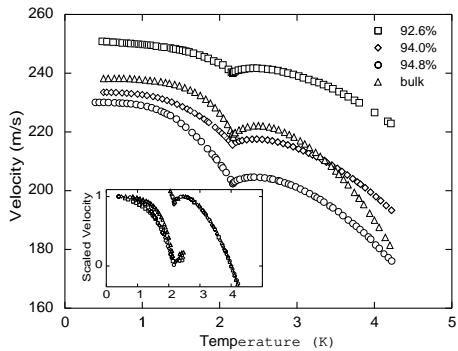}
\caption{\label{velocity} Sound velocity for various aerogels as a function of temperature.
That of bulk helium is also shown for comparison.
The inset shows the scaled sound velocity.
Velocity is scaled between 2.5 and 4.2 K, and 0.5 K and $T_c$, in the normal and the superfluid phase (see in the text).
}
\end{figure}

\begin{figure}[htb]
\includegraphics[width=6cm,keepaspectratio]{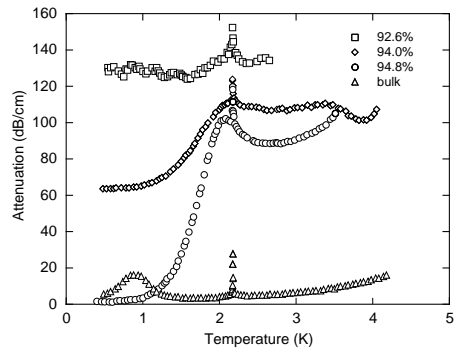}
\caption{\label{attenuation} Attenuation of sound for various aerogels as a function of temperature.
That of bulk helium is also shown for comparison.
For clarity, that for 94 and 92.6 \% is shifted +20 and +40 dB/cm, respectively. 
Those in the normal phase vary due to interference between the ultrasound signal and electrical feed-through from the transducer and amplifier.
}
\end{figure}

Figure~\ref{velocity} and~\ref{attenuation}, respectively, shows the sound velocity $c$ and attenuation $\alpha$ for the three aerogels as a function of temperature;
for comparison, those of bulk helium obtained in a different run were plotted as well.
The superfluid transitions T$_c$ were identified as a dip in velocity and by an absorption peak for each aerogel.
Those in aerogels were as sharp as in bulk, which represented a homogeneous transition in aerogel.
T$_c$ in aerogels is 2.165, 2.168, and 2.168 K for 92.6, 94.0, and 94.8 \% porosity, respectively. 
%are listed in Table~\ref{table1}. 
The magnitude of T$_c$ suppression and porosity dependence agree with specific heat measurement\cite{Wong,Chan}.

%Results velocity%
%
%Results attenuation%
%

%sound velocity%
%\subsection{SOUND VELOCITY}
%
The temperature dependence of the sound velocity is similar to that of bulk for each aerogel.
The absolute value varies inversely to the porosity.
In ordinary porous media, the sound velocity is modified by tortuosity and
the acoustic index $n$ decreases with increasing porosity.
The absolute value varies in opposition to the porosity dependence of the acoustic index; 
thus, tortuosity may not explain the behavior, but may still have an effect.
The observed signal results from a compressional wave within liquid helium modified  by aerogel.

The similarity of the temperature variation brings a scaled behavior to mind. 
Velocity for each aerogel and bulk liquid was scaled as $(c(T)-c(4.2K))/(c(2.5K)-c(4.2K))$ and $(c(T)-c(T_{c}))/(c(0.5K)-c(T_{c}))$ in normal and superfluid phase, respectively. 
The scaled temperature variations in the normal phase for each aerogel and for bulk liquid coincide with each other as shown in the inset of Fig.~\ref{velocity}.
This means the temperature variation is determined mainly by the bulk liquid.
However, that in the superfluid phase depends on the porosity. 
We will discuss this behavior in detail below.
%compare with Vycor system%%
It is useful to compare it with the longitudinal sound velocity in $^4$He filled Vycor\cite{BeamishHikata}.
In the normal phase, the sound velocity in the Vycor system was almost constant, reflecting the constant sound velocity of Vycor glass.
Contrary to the aerogel case, Vycor glass dominated the sound velocity of the composite system. 

%normal phase%
In a series of papers\cite{Biot1,Biot2,Biot3,Biot4}, Biot proposed
a phenomenological theory of acoustic propagation in porous, fluid
filled, macroscopically homogeneous and isotropic media. 
We applied Biot's theory in the low frequency limit since the viscous
penetration depth is larger than the mean separation of SiO$_2$
strands. 
Biot argued that there are two (fast and slow) modes in the composite system. 
The observed signal in the present study corresponds to the fast mode. 
The corresponding mechanical properties of aerogel and the complex system were evaluated by fitting the experimental data for each aerogel. 
We could fit the temperature dependence with this theory only if the coupling constant was nearly zero or
even negative for each aerogel. 
As shown by Johnson\cite{Johnson}, $n$ is given as square root of the coupling constant so that this becomes nearly zero.
According to this, $n$ as evaluated by the Biot theory had no physical meaning.
The Biot theory is applicable to a situation in which sound propagation is mainly determined by a solid and fluid provides a small perturbation to the system and works well in $^4$He filled Vycor\cite{BeamishHikata} and alumina ceramics\cite{WarnerBeamish}.
The equation which represents the sound velocity\cite{Johnson} has no analytical solution in the case that sound velocity of solid and liquid is close.
The aerogel-liquid $^4$He system is really this case.
We concluded that Biot's theory is not applicable to the aerogel-liquid $^4$He case.

Here we present only a very simple phenomenological model for sound
propagation in the normal phase, a more complicated theory will be
published elsewhere\cite{MatsuTayu}. 
On the assumption that two different elastic media (the fluid and aerogel) are in parallel, 
the bulk modulus of the composite medium $K_m$ can
be estimated to be $\phi K_{He} + K_{a} $ using that of the fluid $K_{He}$ and aerogel $K_a$. 
The total density is expressed using the density of aerogel $\rho_a$, that of
liquid helium $\rho_{He}$ and $\phi$ as $\rho_a +
{\phi}\rho_{He}$. Then, the sound velocity, c is expressed as
%equation of velocity in the normal phase%
\begin{equation}
\label{velocitynormal}
c^2 = \frac{\phi K_{a} + K_{He}}{\rho_a + {\phi}\rho_{He}}
= \frac{c_a^2 \rho_a + {\phi}c_{He}^2 \rho_{He}}{\rho_a + {\phi}\rho_{He}} ,
\end{equation}
where $c_a$, $c_{He}$ is the sound velocity of aerogel and that of liquid helium, respectively.

Equation \ref{velocitynormal} has been used to estimate the sound velocity of aerogel by fitting the temperature dependence of the observed sound velocity in the normal phase.
The aerogel sound velocities are assumed to be constant with temperature,
considering other experiments\cite{XieBeamish,DaughtonMulders}.
The best fitting is obtained with the aerogel sound velocity,  
256, 212, and 181 m/s for 92.6, 94.0 and 94.8 \% porosity, respectively. 
%which are listed in Table~\ref{table1}.
These values are consistent with the sound velocity obtained by Gross et al.\cite{Fricke}.
We will use these values to analyze the sound mode in the superfluid phase.

%table of the properties of various aerogels%
%obtained sound velocity %
%\begin{table}[htb]
%\begin{table}[tb]
%\caption{The porosity $\phi$, superfluid transition temperature $T_c$, and sound velocity of aerogel  $c_a$ evaluated from the data in the normal phase.}
%\label{table1}
%\begin{center}
%\begin{tabular}{cccc}
%\hline
%\begin{tabbing}
%{No.}& {$\phi$ (\%)}& {$T_c$ (K)}& {$c_a$ (m/sec)} \\ \hline
%1 & 92.6 & 2.165 & 256 \\
%2 & 94.0 & 2.168 & 212 \\
%3 & 94.8 & 2.168 & 181 \\ \hline
%\end{tabbing}
%\end{tabular}
%\end{center}
%\end{table}
%

\begin{figure}[htb]
\includegraphics[width=6cm,keepaspectratio]{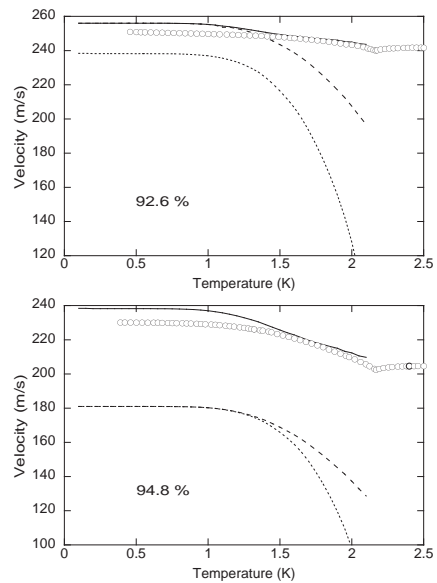}
\caption{\label{velocitycomp} 
Comparison between the observed sound velocity and that of fast mode calculated by the hydrodynamic equations in the superfluid phase. Circles represent experimental points and lines are theoretically calculated.}
\end{figure}

For numerical calculations in the superfluid phase, we use
two-fluid hydrodynamic equations which take into account the
ability of aerogel to move \cite{Brusov}. 
The velocities of three sound modes calculated for 92.6 and 94.8
\% as above are shown in Fig.~\ref{velocitycomp}. 
The solid, broken and dotted line corresponds to fast, intermediate and slow
mode, respectively. 
The sound velocity of these modes converges to that of bulk helium and aerogel, since there is neither normal
component or viscous coupling between the two media at low temperatures. 
The slow mode velocity goes to zero at T$_c$. 
The intermediate acoustic mode includes the complicated motion of all
three components in our system (normal and superfluid components 
and aerogel) and one needs additional investigations of coupling
mechanism between $^4$He and aerogel to get more detailed pictures about
this mode as well as two others.
It is clearly shown that the experimentally observed sound mode corresponds to the fast mode; 
these agree well between 1 K and $T_{c}$ for all aerogels.
However, the discrepancy becomes significant below 1 K. 
The calculated fast mode converges to the greatest sound velocity in
the composite system (for 94.0 and 94.8 \% aerogel - to bulk liquid
velocity, for 92.6 \% aerogel - to aerogel sound velocity). 
On the other hand, the experimentally observed sound velocities at
low temperature are lower than the calculated values. 
The porosity dependence of the velocity could not result from the tortuosity
as in the case of normal phase. 
Then, the coupling between liquid and aerogel should be considered apart from viscosity of the
normal fluid. 
We compared mean free path of phonons and rotons
and that determined geometrically by aerogel strands. 
The geometrically limited mean free path becomes shorter than that of
phonons and rotons below 1K. Acoustic phonons are thought to be
scattered by aerogel strands and to give rise to the momentum
transfer between aerogel and phonons. This means that the simple
hydrodynamic theory is not applicable to this temperature range
because there is no mechanism of momentum transfer due to there being no
viscous fluid. A new theory is necessary in which momentum
transfer between aerogel and phonon should be taken into account
as in the case of the liquid $^3$He-aerogel system\cite{Nagai}. 
We applied the simple idea used in the normal phase to sound velocity at low temperatures 
because the ultrasound wavelength was long enough to regard the micro structure of
aerogel as homogenous and there was only one fluid component. 
The calculated sound velocities using Eq.~\ref{velocitynormal} and aerogel sound velocities obtained by fitting in the normal phase
%listed in Table~\ref{table1} 
are slower than experimental ones for all aerogels; 
this may result from weaker coupling compared to the normal phase. 
A detailed coupling mechanism seems necessary to fit the sound velocity at low temperatures.

%Results velocity comparing with calculated 3 modes and experimental data%
%
%attenuation%
%\subsection{ATTENUATION}
%
The attenuations in the normal phase were so large that the attenuation variations are experimental artifacts resulting from inadvertent interference between the ultrasound signal and electrical feed-through from the transducer and amplifier.
There is no substantial porosity dependence in the normal phase.
At T$_c$, a sharp absorption peak was observed for each aerogel as was observed in the bulk liquid. 
In the specific heat measurement\cite{Wong,Chan}, two distinct singularities of specific heat were observed.
The higher temperature singularities were verified as coincident with the bulk liquid singularities. 
In our experiment, the absorption peak which corresponds to the T$_{\lambda}$ was not observed for every aerogel.
In fact, the temperature resolution of our experiment was not as good as the specific heat measurements\cite{Wong,Chan}, 
but was adequate to distinguish between the peaks at T$_c$ and T$_{\lambda}$.
In the sound experiment, the attenuation due to the bulk liquid may be obscured by the large attenuation of the aerogel.

The observed constant attenuation in the superfluid phase below 1 K
strongly depends on the porosity (or in other words on the density
of SiO$_2$ strands that act as scattering centers) and there is no
temperature dependent contribution from phonons and rotons. 
This behavior qualitatively agrees with the geometrically limited
mean free path picture. 
The general tendency of our results (attenuation is large in the normal phase and decreases with
temperature in the superfluid phase) is similar to those of
alumina ceramics\cite{WarnerBeamish}.

The absorption peak around 1 K which is observed in bulk liquid was not observed in the aerogel system.
In the bulk liquid, the absorption peak is due to the phonon-roton interaction and the peak appears at the temperature that corresponds to $\omega\tau \sim 1$, where $\omega$ is the angular frequency of sound and $\tau$ is the relaxation time.
The absence of the attenuation peak can be qualitatively explained as follows:
The phonon mean free path increases with decreasing temperature and finally exceeds the geometrically limited mean free path in the aerogel.
The relaxation time is limited by the aerogel and cannot satisfy the relation $\omega\tau \sim 1$, so that the attenuation peak is not observed.
Similar peak suppression has been observed in the $^4$He-$^3$He mixture\cite{HardingWilks},
where the existence of $^3$He excitations means that there will be more collisions, or in other words $\tau$ is smaller than that in pure $^4$He.
The peak of the attenuation occurs at lower temperatures compared to pure $^4$He. Because of the lower temperature, the number of phonons and rotons is lower and the value of the peak attenuation is reduced compared to that of bulk $^4$He.
The same scenario may apply in the aerogel-$^4$He system.

%conclusions%
In conclusion, we have studied the low temperature acoustic
properties of a liquid $^4$He filled aerogel system for aerogels of various
porosities and observed a compressional wave in both the
normal and superfluid phases using 10 MHz ultrasound. 
It has been found that sound velocity and attenuation are strongly
influenced by aerogel. 
The scaling behavior has been found in the normal phase. In the
superfluid phase, the three sound modes are calculated from the
hydrodynamic model and the observed sound mode has been shown to
correspond to the fast mode. 
The attenuation peak due to the phonon-roton interaction 
has not been observed in the present  system. The geometrical
limited phonon mean free path by aerogel strand seems to play an 
important role in the acoustic properties at low temperature. 
More detailed theoretical consideration is in progress\cite{MatsuTayu}. 

%Acknowledgement%
%\section{ACKNOWLEDGEMENT}
This work was partially supported by a Grant-in-Aid for Scientific Research from the Ministry of
Education, Culture, Sports, Science and Technology of Japan.
The authors would like to express their appreciation to H. Suzuki and S. Abe for their useful discussions.
They also wish to thank T. Higaki for his contribution to the early stage of this study.

%\bibliography{999}

\end{document}